\documentclass[11pt]{article}

\usepackage{mathrsfs}
\usepackage{bbm,bm}
\usepackage{multirow}
\usepackage{enumerate}
\usepackage{graphicx}
\usepackage{color}
\usepackage{hyperref}
\usepackage{stmaryrd}
\usepackage{mathtools}

\SetSymbolFont{stmry}{bold}{U}{stmry}{m}{n} 

\usepackage{xcolor}
\hypersetup{
    colorlinks,
    linkcolor={red!50!black},
    citecolor={blue!50!black},
    urlcolor={blue!80!black}
}

\usepackage[authoryear]{natbib}

\DeclareMathAlphabet{\mathcalold}{OMS}{cmsy}{m}{n}
\DeclareMathAlphabet{\bmathcalold}{OMS}{cmsy}{b}{n}
\DeclareMathAlphabet{\mathbcal}{OMS}{cmsy}{b}{n}

\usepackage{graphicx,mathtools}
\usepackage{multirow} 
\usepackage{subfigure} 
\graphicspath{{../Graphics/}}
\usepackage{amsmath,amssymb}
\usepackage{hyperref}
\usepackage{color}
\usepackage{tcolorbox}
\tcbuselibrary{xparse}
\usepackage{relsize}
\usepackage{multirow} 
\usepackage{latexsym}
\usepackage{natbib,cuted}
\usepackage{cancel}
\usepackage{array}
\hypersetup{
    colorlinks=true,
    linkcolor=blue,
    filecolor=magenta,      
    urlcolor=cyan,
}
\allowdisplaybreaks

\usepackage{geometry}
\newcommand{\bB}[1]{\boldsymbol{#1}}
\definecolor{ros}{RGB}{148,35,9}   
\usepackage{upgreek}

\def\tsc#1{\csdef{#1}{\textsc{\lowercase{#1}}\xspace}}
\tsc{WGM}
\tsc{QE}

\newcommand{\divr}{{\rm div}\nonscript\mspace{-\muexpr\medmuskip*1/9}}
\newcommand{\surfactant}{${\rm C}_{12}{\rm E}_3$}

\usepackage{cuted} 

\usepackage[font=small,labelfont=bf,margin=12pt]{caption}
\usepackage{multirow}
\newcommand{\rfrf}[1]{??}

\usepackage{wrapfig}

\usepackage{siunitx}

\usepackage{changepage}
\usepackage{pdfpages}

\begin{document}

\title{Application of the thin-film equations in modelling of Marangoni flow patterns amongst surfactant source and drain locations}

\author{
{\sc Julio Careaga}\thanks{Departamento de Matem\'atica, Universidad del B\'io B\'io, Chile, email:{\tt jcareaga@ubiobio.cl}.}
\quad
{\sc Peter A. Korevaar}\thanks{Institute for Molecules and Materials, Radboud University, Heyendaalseweg 135, 6525 AJ Nijmegen, The Netherlands, email:{\tt peter.korevaar@ru.nl}.}
\quad
{\sc Vanja Nikoli\'c}\thanks{Department of Mathematics, Radboud University, Heyendaalseweg 135, 6525 AJ Nijmegen, The Netherlands, email:{\tt v.nikolic@math.ru.nl}.}
\quad
{\sc Laura Scarabosio}\thanks{Department of Mathematics, Radboud University, Heyendaalseweg 135, 6525 AJ Nijmegen, The Netherlands, email:{\tt l.scarabosio@math.ru.nl}.}
}

\date{ }
\maketitle

\begin{abstract}
Surfactants that are deposited at aqueous liquid films have the ability to generate surface tension gradients at the air-water interface, and thereby induce Marangoni flow. Combined with the production and depletion of surfactants at different locations of source and drains, out-of-equilibrium surface tension gradients can be sustained, resulting in Marangoni flow patterns that drive e.g., self-organization of amphiphile myelin assemblies. Here, a mathematical model based on the thin-film equations is proposed to simulate these flow patterns. The model equations are based on the surfactant source and drain concentrations, film-height and surfactant bulk concentration. We present a numerical scheme for approximating the model equations and discuss the numerically observed properties of the model.
\end{abstract}\


\maketitle

\section{Introduction}
\noindent Interfacial flows induced by surfactants, or simply Marangoni flows, appear in a wide range of chemical applications and biological processes. When surfactant molecules are locally produced at the air-water interface and depleted at a different location, an out-of-equilibrium surface tension gradient is sustained which drives a Marangoni flow along the air-water interface \citep{Nguindjel2022}. For example, the rove beetle Microvelia uses this concept to move rapidly over water \citep{Bush2006}, and in synthetic contexts, Marangoni flows have been used as a driving force by maze-solving droplets \citep{Lagzi2010}, in soft-robotics \citep{Yoshida2022} and in self-organization of floating hydrogel structures \citep{Lu2023}.
\begin{figure*}[!t]
\centering
 \includegraphics[scale=1]{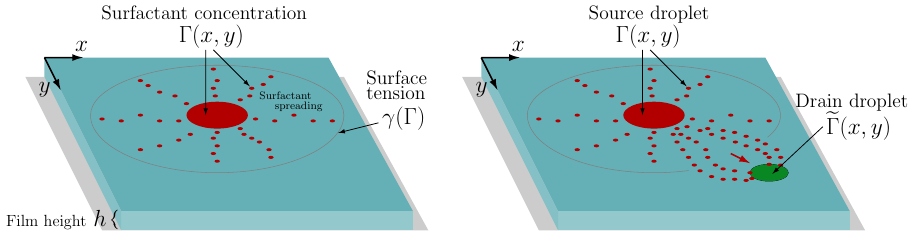}
\caption{(left) Illustration of the spreading of source surfactant of concentration $\Gamma$ on a thin liquid film. (right) Interaction between the source and drain droplet of concentration $\widetilde{\Gamma}$. \label{fig:illustration}}
\end{figure*}
When combined with supramolecular assembly, surfactants involved in the Marangoni flow can also form long filaments, which follow the Marangoni flow and thereby result into self-organized structures. Earlier, it has been shown how myelin filaments spontaneously grow from source droplets of amphiphiles deposited at an air-water interface. Concomitantly, these source droplets release individual amphiphile molecules as surfactant to the air-water interface. As these surfactants are depleted by drain droplets present at the air-water interface as well, a Marangoni flow is established from the source towards the drain droplets, such that the myelin filaments grow from the sources to the drains -- as they follow the Marangoni flows \citep{Nguindjel2021, Nguindjel2022, Winkens2022, Winkens2023}. The dynamics from such systems of surfactants, which rely on the interplay of self-assembly, surfactant release and depletion, and Marangoni flow, can be harnessed to simulate materials with life-like properties.

From a mathematical modeling perspective, Marangoni flows induced by surfactants have been widely studied by making use of Lubrication theory \citep{Greenspan1978, Oron1997, AfsarSiddiqui2003}. After assuming that the height of the liquid film is sufficiently small with respect to its characteristic horizontal length, the transport and Navier-Stokes equations, written for the concentration of surfactant and flow velocity, are reduced to the so-called \emph{thin-film equations}. These equations allow to model the surfactant spreading and liquid height (free surface) by nonlinear, high-order and time dependent partial differential equations, which depend mainly on the surface tension function of the liquid. Approaches based on thin-film equations have the main advantage of reducing the complexity of the phenomenon by one spatial dimension, avoiding determining three-dimensional vector field flows. Several applications of different kind can be found in the literature. \cite{Borgas1988} studied a steady-state version of the thin-film equations for the modeling of lung lining flows. They analyze surface diffusion and spreading rates depending on the model parameters.
A transient model for the inhaled aerosol on the lung's liquid lining is proposed by \cite{Gaver1990}. \cite{Jensen1993} incorporate the bulk concentration on the liquid as an additional unknown, coupling a convection diffusion equation to the thin-film equations. In addition, they consider variations due to temperature fluctuations, and adsorptive and desorptive rates are included as source terms into the equations.
The model by \cite{Eres1999} assumes that the surface tension depends solely on the bulk concentration, excluding the unknown related to the surface concentration from the model. They also include extra ingredients such as a nonlinear concentration dependent viscosity function, a diffusion coefficient, and the solvent evaporation.
\cite{Matar1999} dedicated their work to the study of fingering formation created by the spreading of insoluble surfactant on the surface of a thin liquid film including van der Waals forces. Performing linear stability analysis, they were able to describe the evolution of  disturbances. Furthermore, two-dimensional simulations are presented.
The work by \cite{Matar2001} extends the thin-film equations to simulate the process of cleaning semiconductors. This model considers three types of concentrations of alcohol, at the bulk, air-water surface and air.

Due to the intricacy of the terms contained in the thin-film equations, no exact solutions are known and their approximation through numerical schemes is a topic on itself. Several numerical methods have been proposed to deal with these equations with their own pros and cons. These include methods based on finite differences \cite{Peterson2012}, finite volumes \cite{Grun2000} and even finite elements \citep{Barrett2003,Barrett2006,Barrett2008,Liu2019}.

In this work, we elaborate a mathematical model for the spreading of amphiphile surfactant under the presence of source and drain droplets based on the thin-film equations. The model is grounded on the theory of continuum mechanics and the surfactant concentration is treated as a single phase, where no filaments can be distinguished. However, it is assumed that the myelin filaments that are assembled by the surfactant are transported following the lines of flow. Furthermore, we propose a numerical scheme to numerically approximate the equations, based on the so-called mixed finite element method. In this approach, we compute two additional horizontal velocities, one for the concentration of surfactant, and one for the wave generated at the water-air interface.

\section{Mathematical model}
\subsection{Description of the process}

We begin by considering a viscous fluid confined to a container or vessel, which has a free boundary at the liquid-air interface (top boundary), and that is initially at rest with constant height $h_0$. In addition, we assume that the vessel has a constant horizontal cross-section $\Omega$, which serves as a two-dimensional domain in our approach. For the special case of a Petri dish, $\Omega$ becomes a circle of radius $R$. The horizontal coordinates are then denoted by $x$ and $y$, and $\bB{x} = (x,y)$ is a point in $\Omega$, while the vertical coordinate is represented by the letter $z$.
The height of the liquid film, which is measured as the distance from the (flat) bottom of the vessel at $z=0\,\rm m$ to the water-air interface (free boundary), is determined by a function $h$ which can vary with respect to the spatial coordinate $\bB{x}\in \Omega$ and time $t>0$, i.e., $h:= h(\bB{x},t)$.
At the beginning of the process, one or several droplets of surfactant (source droplets) at a certain concentration are deposited into the water-air interface. The concentration of surfactant at this interface is denoted by $\Gamma:=\Gamma(\bB{x},t)$, which is a function depending on $\bB{x}\in \Omega$ and $t$. The surfactant is then modeled as a continuous medium, whose concentration can be determined locally on each point in $\Omega$. Right after $t=0\,\rm s$, due to changes in the surface tension of the liquid film produced by the source droplets, the liquid film breaks the equilibrium, and the particles of surfactant begin to spread through the water surface.
We consider source droplets of circular shape, therefore it is expected that initially the particles will be transported along a radial direction.
Furthermore, we are going to assume that part of the surfactant begins to mix into the liquid. This bulk concentration of surfactant diluted in the liquid will be denoted by $c$, which is assumed to be a function varying only in the horizontal coordinates and time, this is, $c:=c(\bB{x},t)$. Then, the concentration $c$ is assumed to be constant along the vertical coordinate.
In the examples shown in Section~\ref{sec:simulations}, we use triethylene glycol monododecylether (\surfactant) as source surfactant.


An essential characteristic of the processes studied by \cite{Winkens2023} is the presence of the so-called drain-droplets, which are droplets of a chemical compound that act as attractors of surfactant. These drain droplets, due to their chemical composition, remain floating at the water-air interface, while the attracted surfactant particles reaching them are getting depleted. Then, in our approach, we assume that at $t=0\,\rm s$, drain droplets of circular shape with a concentration $\widetilde{\Gamma}:=\widetilde{\Gamma}(\bB{x},t)$ are located at the water-air interface.

\begin{figure*}
\centering
 \includegraphics[scale=0.5]{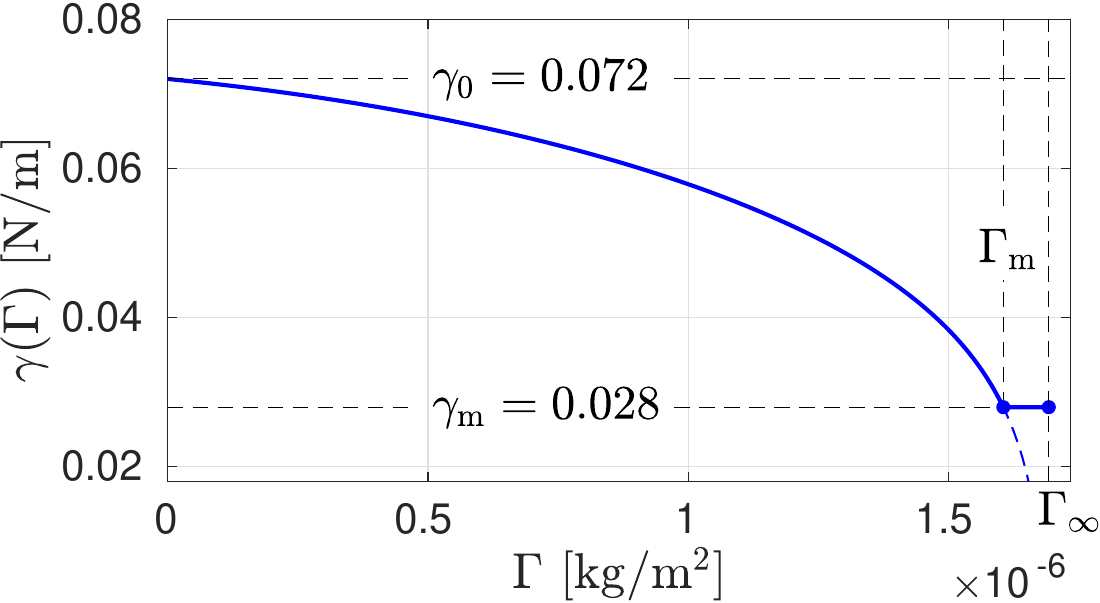}
 \caption{Surface tension function $\gamma$ given by formula \eqref{eq:gamma:frumkin} with constants $R = 22.904856\,\rm J/kg\,C_{12}E_{4}K$, $T = 298.15\, \rm K$, and $K_{\rm a} = 1.875$, and truncated within the interval $\Gamma_{\rm m}<\Gamma \leq \Gamma_{\infty}$. The blue dashed line continuing from the point $(\Gamma_{\rm m},\gamma_{\rm m})$ corresponds to the path of the non-truncated function.
 Here $\gamma_0 = 0.072\,\rm N/m$ is the surface tension of pure water, $\gamma_{\rm m} = 0.028\,\rm N/m$ is the micelle surface tension, where $\Gamma_{\rm m} = 1.605235804\times10^{-6}\,\rm kg/m^2$ and $\Gamma_{\infty} = 1.69266\times10^{-6}\,\rm kg/m^2$.}\label{fig:plotgamma}
\end{figure*}

The primary component that induces the Marangoni flow is the difference in surface tension produced by the existence of surfactant at the water-air interface. This is modeled by the surface tension function $\gamma$, which is assumed to only depend on the concentration of the surfactant, i.e., $\gamma:=\gamma(\Gamma)$. One of the main characteristics of the surface tension function is that it is decreasing with respect to $\Gamma$. The maximum surface tension value is denoted by $\gamma_0$ and corresponds to the case of pure water or zero surfactant concentration, so we impose $\gamma(0) = \gamma_0$.  On the other hand, its minimum value is assumed when the surface concentration reaches $\Gamma_\infty$, the maximum surface concentration. Several surface tension functions have been proposed in the literature, from linear, quadratic or power law relations, to decreasing exponential functions \citep{Hanyak2012}. In this work, we use the Frumkin adsorption isotherm proposed by \citep{Hsu2000}. For $\Gamma < \Gamma_{\rm m}< \Gamma_{\infty}$, we set
\begin{align}\label{eq:gamma:frumkin}
 \hspace{-0.5cm}\gamma(\Gamma) := \gamma_0 + \Gamma_{\infty}RT \left({\rm ln}\left(1 - \dfrac{\Gamma}{\Gamma_{\infty}}\right) - \dfrac{K_{\rm a}}{2} \dfrac{\Gamma^2}{\Gamma_{\infty}^2}\right),
\end{align}
and for $\Gamma>\Gamma_{\rm m}$, we define $\gamma(\Gamma) = \gamma(\Gamma_{\rm m}) = \gamma_{\rm m}$, which corresponds to the value of surface tension when the surface is full of surfactant, at the critical micelle concentration $\Gamma_{\rm m}$. Here $R$ is the molar gas constant, $T$ is the temperature, and $K_{\rm a}$ is a constant related to the adsorption rate. These constants have the following values
\begin{align*}
 R & = 22.904856\,\rm J/kg\,C_{12}E_{4}K,\\
 T & = 298.15\, \rm K,\\
 K_{\rm a} & = 1.875.
\end{align*}
Figure~\ref{fig:plotgamma} shows the plot of function $\gamma$ for values of $\Gamma$ between $0$ and $\Gamma_{\infty}$.
Observe that, since the definition of $\gamma$ involves the logarithm term which tends to infinity when its argument tends to zero, the value of $\Gamma_{\rm m}$ must be strictly less than $\Gamma_{\infty}$. Finally, Figure~\ref{fig:illustration} shows the spreading of $\Gamma$ over the surface of a thin liquid film of height $h$, and how this is meant to interact with the drain droplet of concentration $\widetilde{\Gamma}$. The wave generated by the gradient of surface tension is illustrated by the red circle.


\subsection{Governing equations}

As described in the previous section, we are interested in determining how the surface surfactant $\Gamma$, height $h$, and bulk surfactant concentration $c$ evolve in space and time as the physicochemical process is carried out. For this purpose, we develop a mathematical model based on Lubrication theory \citep{Greenspan1978}, i.e. the problem is reduced by one space dimension and the concentration of surfactant is modeled as a continuum media. Therefore this approach does not cover the formation of filaments as individual identities.

We first observe that $\Omega$ is a two-dimensional bounded domain in Cartesian coordinates, and define its characteristic length as $L$, which in case of a circular domain corresponds to its diameter.
Typically, solid-liquid two-phase flow problems are modeled through the Navier-Stokes momentum equations (NSEs) for the flow field \citep{Ungarish1993,Crowe2011,Ishii2006}, and equations of mass conservation equation (MCE). In the case of the surface surfactant, these equations should be written at the moving surface $z=h(\bB{x},t)$. The thin-film equations are derived from the three-dimensional NSEs and MCE for the surface surfactant after assuming that the height of the liquid film is sufficiently smaller than $L$ \citep{Oron1997,Hanyak2012}. In this approach, the horizontal velocity field $\bB{v}\in\mathbb{R}^2$, related to the fluid phase at a point $(\bB{x},z)$ in $\mathbb{R}^3$, and the  pressure $p$ at the interface $z=h(\bB{x},t)$ are given by
\begin{align*}
 \bB{v}(\bB{x},z,t) 
       & = \dfrac{1}{\mu} \left(\dfrac{z^2}{2}-hz\right) \nabla p  + z\nabla\gamma(\Gamma),\\
   p(\bB{x},h,t) & = \rho g h - \gamma(\Gamma) \nabla^2h,
\end{align*}
where $g$ is the acceleration of gravity, $\rho$ is the fluid density, $\mu$ is the viscosity of the fluid, and $\gamma$ is the surface tension function defined in \eqref{eq:gamma:frumkin}.
Here, the gradient, divergence and Laplacian operators, denoted by $\nabla$, $\divr$ and $\nabla^2$, respectively, are defined with respect to the horizontal coordinates $x$ and $y$. The velocity field at the boundary layer $z=h(\bB{x},t)$, which drives the spreading of surfactant through the water surface, is then determined by
\begin{align*}
\bB{v}_{\rm s} := \bB{v}_{\rm s}(\bB{x},t):= \bB{v}(\bB{x},h(\bB{x},t),t),
\end{align*}
with subscript '$\rm s$' referring to the word 'surface'. Moreover, we define $\bB{w}:=\bB{w}(\bB{x},t)$ as the two-dimensional vector corresponding to the following average velocity over the vertical component
\begin{align*}
 \bB{w} = \dfrac{1}{h}\int_0^h\bB{v}(\bB{x},z,t){\rm d}z,\quad\text{for }h>0.
\end{align*}
The definitions of $p$, $\bB{v}_{\rm s}$ and $\bB{w}$ together with equations for the conservation of mass and volume provide a set of equations for the evolution of the film-height $h$, surfactant concentration $\Gamma$, and bulk concentration $c$. Then, the governing equations described for each point $\bB{x}$ in $\Omega$ and time $t>0$ are
\begin{subequations}\label{syst:main:evolution}
\begin{align}
 \partial_t h + \divr(h\bB{w})& = 0,\label{eq:mainevol:a}\\
 \partial_t \Gamma + \divr(\Gamma\bB{v}_{\rm s}) & = \mathcal{D}_{\rm s}\nabla^2\Gamma + \mathcal{R}_{\rm s}(\widetilde{\Gamma},\Gamma,c) + \psi(\Gamma),\label{eq:mainevol:b}\\
 \partial_t c + \bB{w}\cdot \nabla c &= \mathcal{D}_{\rm b}\nabla^2 c + \mathcal{R}_{\rm b}(\Gamma,c), \label{eq:mainevol:c}
 \end{align}
\end{subequations}
where $\mathcal{R}_{\rm s}$ and $\mathcal{R}_{\rm b}$ are reaction terms related to the surfactant and bulk, respectively, $\psi=\psi(\Gamma)$ is a source term, and $\mathcal{D}_{\rm s}$ and $\mathcal{D}_{\rm b}$ are diffusion coefficients. The velocities $\bB{v}_{\rm s}$ and $\bB{w}$, and pressure $p$ are determined by
\begin{subequations}\label{syst:main:others}
\begin{align}
 \bB{v}_{\rm s} &= \dfrac{1}{\mu} \left(h\nabla\gamma(\Gamma) -\dfrac{1}{2}h^2\nabla p\right), \label{eq:maino:a}\\
 \bB{w} &= \dfrac{1}{\mu} \left(\dfrac{1}{2}h\nabla\gamma(\Gamma) -\dfrac{1}{3}h^2\nabla p\right),\label{eq:maino:b}\\
 p &= \rho g h - \gamma(\Gamma)\nabla^2 h.\label{eq:maino:c}
\end{align}
\end{subequations}
In the above system of equations, capillary effects are already incorporated through the pressure gradient $\nabla p_{\rm cap} = -\nabla(\gamma(\Gamma)\nabla^2 h)$, and the Marangoni flow is included into the equations \eqref{eq:maino:a} and \eqref{eq:maino:b} via the term $\nabla\gamma(\Gamma)$. Since the surface tension decreases as $\Gamma$ increases, $\gamma'(\Gamma)<0$ for all $\Gamma$.
Note that the concentration corresponding to the drain droplet $\widetilde{\Gamma}$ is not being determined by any partial differential equation in the system. In fact, this concentration is assumed to be given as a known function, which will mainly serve to determine the region in which the drain droplet is located. More precisely, the effect of the drain droplet is going to be modeled as a sink term in the equations.
The source term $\psi(\Gamma)$ is incorporated in equation \eqref{eq:mainevol:b} in order to model the continuous release of surfactant made by the source droplet. 

The system \eqref{syst:main:evolution}--\eqref{syst:main:others} needs to be supplemented with suitable boundary and initial conditions. In order to have zero flux at the boundary $\partial\Omega$, we impose the following boundary conditions:
\begin{align*}
 \bB{n}\cdot \bB{v}_{\rm s} = 0, \quad \bB{n}\cdot \bB{w} = 0,\quad  \bB{n}\cdot \nabla h = 0&\quad \mbox{on }\partial\Omega, \\
 \bB{n}\cdot \nabla \Gamma = 0,\quad \bB{n}\cdot \nabla c = 0&\quad \mbox{on }\partial\Omega.
\end{align*}
For the initial conditions (variables at $t=0$), we set:
\begin{align*}
 h(\bB{x},0) = h_0, \quad  \Gamma(\bB{x},0) = \Gamma_0(\bB{x}),\quad c(\bB{x},0) = 0,
\end{align*}
for $\bB{x}\in\Omega$, where $h_0>0$ is constant and $\Gamma_0(\bB{x})$ corresponds to the concentration of surfactant making up the source droplets distributed in circular regions within the domain $\Omega$.

Finally, we remark that by substituting \eqref{eq:maino:c} into \eqref{eq:maino:b} and consecutively \eqref{eq:maino:b} into \eqref{eq:mainevol:a} one recovers a fourth-order nonlinear degenerate partial differential equation (PDE) in terms of the height $h$. This indicates that the analysis and simulation of \eqref{syst:main:evolution}--\eqref{syst:main:others} is particularly challenging.
\subsection{Source and drain droplets}

In this work, we assume that all droplets have circular shapes. To properly define them, we introduce the notation $\mathcal{C}(\bB{x}_{0},r_0)$ to refer to the set of points in $\Omega$ located within the circle of radius $r_0$ and center $\bB{x}_0$ in $\Omega$.

For the source droplets, we assume that $\Gamma_0$ is composed by $N$ circular disjoint droplets strictly contained in $\Omega$, each one of radius $r_i$ and center point $\bB{x}_i$, for $i=1,2,\dots,N$. For each circular droplet $i=1,2,\dots,N$, and $\bB{x}$ in $\mathcal{C}(\bB{x}_i,r_i)$, we define the concentration $\Gamma_0$ as follows. On an inner circle we set
\begin{align*}
 \Gamma_0(\bB{x}) = \Gamma_i,\qquad \text{ for $\bB{x}$ in $\mathcal{C}(\bB{x_i},\alpha r_i)$},
\end{align*}
for $0<\alpha<1$, where $\Gamma_i>0$ is the characteristic concentration of the droplet. For the remaining part of $\mathcal{C}(\bB{x}_i,r_i)$, the radius comprehended between $\alpha r_i$ and $r_i$, we set
\begin{align*}
 \Gamma_0(\bB{x}) = \dfrac{\Gamma_i}{r_i^2(1-\alpha^2)} (r_i^2-r^2),\quad \text{ for $\alpha r_i<r<r_i$},
\end{align*}
where $r$ is the radius (or distance) measured from the point $\bB{x}$ to the center $\bB{x}_i$. The above definition is made in order to produce a continuous transition between the value $\Gamma_i$ and the zero concentration outside the circle $\mathcal{C}(\bB{x}_i,r_i)$. In regions of $\Omega$ not covered by any of the circular droplets, we assume zero concentration, this is
\begin{align*}
 \Gamma_0(\bB{x})=0,\quad  \text{for $\bB{x}$ not in }\mathcal{C}(\bB{x}_1,r_1)\cup\dots \cup\mathcal{C}(\bB{x}_N,r_N).
\end{align*}
In a similar fashion, for the drain droplets, we assume that there exist $\widetilde{N}$ disjoint circular droplets strictly contained in $\Omega$, each one of radius $\widetilde{r}_i$ and center $\widetilde{\bB{x}}_i$ in $\Omega$, given by $\mathcal{C}(\widetilde{x}_i,\widetilde{r}_i)$ for $i=1,\dots,\widetilde{N}$. Then, for each circular drain droplet $i=1,\dots,\widetilde{N}$, we set
\begin{align*}
 \widetilde{\Gamma}(\bB{x},t) = \widetilde{\Gamma}_i,\quad \text{for $\bB{x}$ in }\mathcal{C}(\widetilde{x}_i,\widetilde{r}_i),
\end{align*}
with $\widetilde{\Gamma}_i>0$ the characteristic concentration of the droplet $i$. Outside the circles $\mathcal{C}(\widetilde{x}_i,\widetilde{r}_i)$ for $i=1,\dots,\widetilde{N}$, we set zero concentration of drain droplet $\widetilde{\Gamma}(\bB{x},t) = 0$. Although we define $\widetilde{\Gamma}$ as a time dependent function, we are considering these droplets to be at a fixed position.

We assume that each of the drain droplets deplete the surfactant at a constant rate $\kappa_{{\rm d},i}>0$ for $i=1,\dots,\widetilde{N}$. In addition, part of the surface surfactant is being transferred to the bulk concentration at a certain rate $\kappa_{\rm b}$. The resulting reaction term in \eqref{eq:mainevol:b} is given by
\begin{align*}
 \mathcal{R}_{\rm s}(\widetilde{\Gamma},\Gamma,c)& = -\big(\kappa_{\rm b} + \chi({\widetilde{\Gamma}})\big)\Gamma,
 \end{align*}
where
\begin{align}\label{eq:def:drain}
\chi(\widetilde{\Gamma}) :=
 \begin{cases}
 \kappa_{{\rm d},i} & \mbox{ if }\widetilde{\Gamma}_i>0\mbox{ for }i=1,\dots,\widetilde{N},\\
0  & \mbox{otherwise}.
\end{cases}
\end{align}
The surfactant transferred from the surface to the bulk concentration $c$ is modeled via the reaction term
\begin{align*}
 \mathcal{R}_{\rm b}(\Gamma,c) &= \kappa_{\rm b}\Gamma.
\end{align*}
We remark that the treatment given to the initial conditions, and numerical approximations later on, are not compatible with a sharp interface model, in which the variables may present discontinuous fronts. However, since the action of the drain droplet is due to a piecewise discontinuous function, this variable is  modeled in a sharp interface fashion.
Lastly, the source term $\psi$ in Equation~\eqref{eq:mainevol:b} models the release of surfactant by the source droplet(s). This is due to the fact that the concentration of the source droplets can be larger than $\Gamma_{\rm m}$. In this case, the surfactant particles are distributed forming a volume on top of the water-air surface, which releases surfactant particles as they spread on the surface. Then, at each source droplet $i$, the source term is defined as
\begin{align*}
 \psi(\Gamma(\bB{x})) = \kappa_{\rm s}(\Gamma_{\rm m}-\Gamma(\bB{x})),\quad \mbox{ if $\bB{x}$ is in $\mathcal{C}(\bB{x}_i,\alpha r_i)$},
\end{align*}
for some $i=1,2,\dots,N$, and $\psi = 0$ otherwise.

\section{Numerical method}

The system \eqref{syst:main:evolution}--\eqref{syst:main:others} includes both spatial and temporal partial derivatives.
To treat the spatial derivatives, we employ a mixed finite element method (MFEM),
in which, at each time step, the six unknowns $h,\Gamma,c,\bB{v}_{\rm s}, \bB{w}$ and $p$ are solved as one single vector of unknowns in a product space.
In addition, finite element methods are based on so-called weak formulations of a PDE, which typically allow lowering the order of the spatial derivatives involved. In our case, we are going to derive a weak formulation involving first order spatial derivatives only. For the time derivatives in \eqref{eq:mainevol:a}--\eqref{eq:mainevol:c}, we are going to use the backward Euler method. Since \eqref{syst:main:evolution} and \eqref{syst:main:others} contain several nonlinear terms, such as $\nabla\gamma(\Gamma)$ and $h^2$, for example, we use the Newton nonlinear solver together with the MFEM formulation on each time iteration.

We now begin with the discretization of our model equations. We let $\Delta t>0$ be the time step and $t^n = n\Delta t$ with $n\in\mathbb{N}$ the time points, therefore the time dependency of the discrete unknowns will be denoted by the superscript $n$, this is $h^n=h(\bB{x},t^n)$ for example. The backward Euler approximation of the time derivative of $h$ reads as follows
\begin{align*}
 \partial_t h(\bB{x},t^n) \approx \dfrac{h^n - h^{n-1}}{\Delta t},\qquad \text{for all $n\in \mathbb{N}$},
%
\end{align*}
to be equated to the right-hand side of \eqref{eq:mainevol:a} at $t=t^n$, and analogously for the time derivatives of $\Gamma$ and $c$.

To introduce the MFEM, we define the set $\mathcal{T}_k$ as a regular triangulation of $\Omega$ made up of triangles $K\in\mathcal{T}_{k}$ where $k$ represents the largest diameter of all the triangles in $\mathcal{T}_k$.
Then, given an integer $\ell\geq0$, we define the space of piecewise polynomial functions of degree less or equal to $\ell$ over $\mathcal{T}_k$ by $\mathcal{P}_\ell (\Omega)$. This means that each function in $\mathcal{P}_\ell (\Omega)$ when restricted to a triangle in $\mathcal{T}_k$ becomes a polynomial of degree less or equal to $\ell$. In addition, we define the following finite element spaces:
\begin{align*}
 \mathrm{Q} := {\rm C}(\bar{\Omega})\cap \mathcal{P}_\ell (\Omega), \qquad \mathbf{H} := \mathrm{Q}\times \mathrm{Q},
\end{align*}
where ${\rm C}(\bar{\Omega})$ is the set of continuous functions from $\bar{\Omega}$ to $\mathbb{R}$.
The set $\rm Q$ is called the Lagrange finite element space, and it is used to numerically approximate continuous solutions in space. On the other hand, the set $\mathbf{H}$ is a vector space, meaning that its elements are vectors in $\mathbb{R}^2$, where each component is a continuous, piecewise polynomial function in space. The scalar unknowns $\Gamma$, $h$ and $c$ will be searched in $\mathrm{Q}$, while the vector unknowns $\bB{v}_{\rm s}$ and $\bB{w}$ in $\mathbf{H}$.

To obtain a weak formulation for the system \eqref{syst:main:evolution}--\eqref{syst:main:others}, we need to multiply each equation by test functions in $\mathrm{Q}$, respectively in $\mathbf{H}$, and then integrate by parts the resulting equations over $\Omega$. In this way, the discrete weak formulation of \eqref{syst:main:evolution}--\eqref{syst:main:others} reads as follows: Find $(h_k^n,\Gamma^n_k,c^n_k, \bB{v}_{{\rm s},k}^n,\bB{w}_k^n,p_k^n)\in \mathrm{Q}^3\times\mathbf{H}^2\times\mathrm{Q}$ such that
\begin{subequations}\label{syst:discrete:evolution}
\begin{gather}
 \begin{split} \label{eq:maind:a}
\hspace{-0.3cm}&\dfrac{1}{\Delta t} \int_{\Omega}(h^{n}_k-h^{n-1}_k)\xi\,{\rm d}\bB{x} - \int_{\Omega} h^{n}_{k}\bB{w}_{k}^{n}\cdot\nabla\xi\,{\rm d}\bB{x}\\
\hspace{-0.3cm}&\quad \,=\, -\int_{\Omega}\mathcal{D}_{\rm h}\nabla h_k^n\cdot\nabla\xi\,{\rm d}\bB{x},\\[-2ex]
 \end{split}
\end{gather}
\begin{gather}
 \begin{split}\label{eq:maind:b}
\hspace{-0.3cm}& \dfrac{1}{\Delta t} \int_{\Omega}(\Gamma^{n}_k-\Gamma^{n-1}_k)\zeta\,{\rm d}\bB{x} - \int_{\Omega} \Gamma^{n}_{k}\bB{v}_{{\rm s},k}^{n}\cdot\nabla\zeta\,{\rm d}\bB{x}\\
\hspace{-0.3cm}&\quad\,=\,-\int_{\Omega}\mathcal{D}_{\rm s}\nabla\Gamma_k^{n}\cdot\nabla\zeta\,{\rm d}\bB{x}  +  \int_{\Omega}(\mathcal{R}_{\rm s}^{n}+\psi^n)\zeta\,{\rm d}\bB{x},\\[-2ex]
 \end{split}
\end{gather}
\begin{gather}
 \begin{split}\label{eq:maind:c}
\hspace{-0.3cm}&\dfrac{1}{\Delta t} \int_{\Omega}(c^{n}_k-c^{n-1}_k)\vartheta\,{\rm d}\bB{x} + \int_{\Omega}\bB{w}_k^{n}\cdot\nabla c\vartheta\,{\rm d}\bB{x}\\
\hspace{-0.3cm}&\quad\,=\, -\int_{\Omega}\mathcal{D}_{\rm b}\nabla c^{n}_k\cdot\nabla\vartheta\,{\rm d}\bB{x} + \int_{\Omega}\mathcal{R}_{\rm b}^{n}\vartheta\,{\rm d}\bB{x},
 \end{split}
\end{gather}
for each test functions $(\xi,\zeta,\vartheta)\in {\rm Q}^3$ and all $n$, where the discretized reaction and source terms are
\begin{align*}
& \mathcal{R}_{\rm s}^{n}:= \mathcal{R}_{\rm s} \bigl(\widetilde{\Gamma},\Gamma^n_k,c^n_k\bigr),\quad
  \mathcal{R}_{\rm b}^{n}:= \mathcal{R}_{\rm b} \bigl(\Gamma^n_k,c^n_k\bigr),\\
 &\psi^n:=\psi(\Gamma_k^n).
\end{align*}
The diffusion term on the right-hand side of Equation \eqref{eq:maind:a}, where $\mathcal{D}_{\rm h}$ is a small positive number, has been included for numerical stability in order to avoid spurious oscillations that can generate negative solutions. Equations \eqref{eq:maind:a}--\eqref{eq:maind:c} are supplemented with the projection of the initial conditions $h_0$, $\Gamma_0$ and $c_0$ on the continuous piecewise polynomial space $\mathrm{Q}$. System \eqref{eq:maind:a}--\eqref{eq:maind:c} is coupled to the following equations:
\begin{align}
 \hspace{-0.7cm}\int_{\Omega} \bB{v}_{\rm s,k}^{n}\bB{\eta}\,{\rm d}\bB{x} &= \dfrac{1}{\mu} \int_{\Omega}\left(h_k^{n}\nabla\gamma_k^n -\dfrac{1}{2}(h_k^{n})^2\nabla p_k^{n}\right)\cdot\bB{\eta}\,{\rm d}\bB{x}, \label{eq:mainod:a}\\
\hspace{-0.7cm}\int_{\Omega} \bB{w}_k^{n}\bB{s}\,{\rm d}\bB{x} &= \dfrac{1}{\mu} \int_{\Omega}\left(\dfrac{h_k^{n}}{2}\nabla\gamma_k^{n} -\dfrac{1}{3}(h_k^{n})^2\nabla p_k^{n}\right)\cdot\bB{s}\,{\rm d}\bB{x},\label{eq:mainod:db}\\
\hspace{-0.7cm} \int_{\Omega}p_k^{n} q\,{\rm d}\bB{x}&= \int_{\Omega} \rho g h_k^{n}q \,{\rm d}\bB{x}+  \int_{\Omega}\nabla h_k^{n}\cdot\nabla(\gamma_k^nq) \,{\rm d}\bB{x},\label{eq:mainod:c}
\end{align}
\end{subequations}
for all test functions $(\bB{\eta},\bB{s},q)\in \mathbf{H}^2\times \mathrm{Q}$ and all $n$, where $\gamma_k^n:=\gamma(\Gamma_k^n)$. The MFEM given by the coupled system \eqref{syst:discrete:evolution} describes a nonlinear system of equations at each time iteration $n$. These nonlinear equations are solved by the Newton-Raphson method.

\section{Simulations}\label{sec:simulations}
For the numerical simulations, we implemented the marching formulas described by system \eqref{syst:discrete:evolution} in the open source finite element library FEniCS \citep{FEniCS}. For the nonlinear solver, we use the Newton-Raphson algorithm provided by this library with null initial guess, absolute and relative tolerance of $10^{-12}$, and the solutions of tangent linear systems at each iteration of the nonlinear solver
resulting from the linearization are made by the multifrontal massively parallel sparse direct solver MUMPS \citep{Amestoy2000}. For all simulations presented in this section, we consider a Petri dish as the container of the liquid film, therefore the two-dimensional domain $\Omega$ is given by a circle of radius $R=1.75\,\rm cm$, whose center is located at the origin $(0,0)$. For the surface tension of clear water we set $\gamma_0 = 0.072\,\rm  N/m$, and for the minimum surface tension, we use $\gamma_{\rm m}  = 0.028\,\rm N/m$. The value of $\Gamma_{\infty}$ corresponding to the \surfactant surfactant taken for the Frumkin function \eqref{eq:gamma:frumkin} is $4.5\times 10^{-10}\, \rm mol/cm^2$ \citep[Table 1]{Hsu2000}, which can be readily converted into $\rm kg/m^3$ units by using the molecular weight of \surfactant ($0.363\,\rm kg/mol$), this is
\begin{align*}
 \Gamma_{\infty} = 1.69266\times 10^{-6}\,\rm kg/m^2.
\end{align*}
With $\Gamma_{\infty}$ being defined, we set the micelle concentration as $\Gamma_{\rm m}=1.60524\times 10^{-6}\,\rm kg/m^2$. Note that $\gamma(\Gamma_{\rm m}) = 0.028\,\rm N/m$.
Some of the constant used in our model are
\begin{alignat*}{3}
 &\rho = 9.81\,\rm m/s^2, \quad &&\mu  = 0.0058\,\rm Pa\,s, \quad &&\rho = 1000\,\rm kg/m^3,\\
 &\mathcal{D}_{\rm h} = 10^{-6},\quad  &&\mathcal{D}_{\rm s} = 10^{-6}, \quad &&\mathcal{D}_{\rm b} = 10^{-6},
\end{alignat*}
and for the time step we choose $\Delta t = 10\,\rm s$.
For each source droplet shown in the simulations below, we set their characteristic concentration as $\Gamma_{\rm m}$, and for the flat inner region of the droplets we use $\alpha = 0.8$. We set the thin-layer initial height to $h_0(\bB{x})\equiv H = 10^{-6}\,\rm m$.
Regarding the mesh and polynomial degree chosen in all examples, we set the meshsize to $k = 0.668837,\rm mm$, resulting in a total of $10750$ triangles, and $\ell = 1$ (piecewise linear polynomials).

In what follows, we are going to denote the examples depending on the number of source and drain droplets by $N$S$\widetilde{N}$D. We simulate the following configurations: 1S1D, 1S4D and 2S1D with moving drain droplet.


\begin{figure*}[!t]
\centering
 \includegraphics[scale=1]{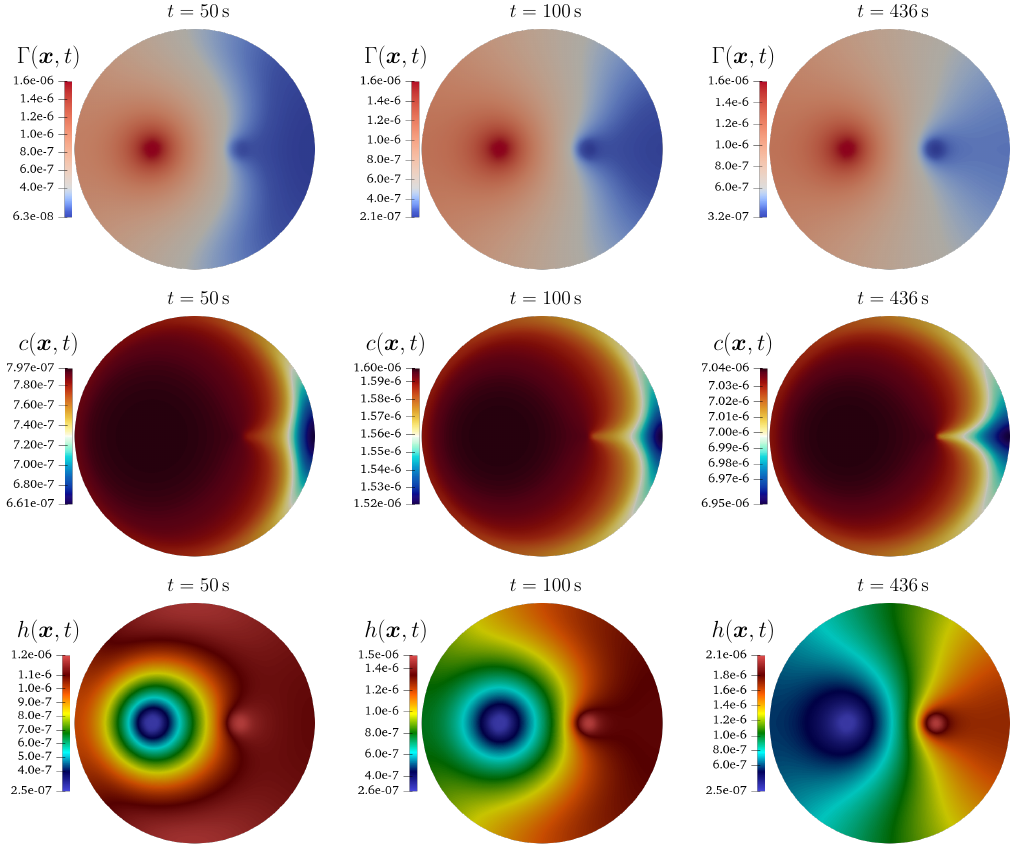}
  \caption{Example 1. Surfactant concentration $\Gamma$ (first row), bulk concentration $c$ (second row) and height $h$ (third row) at $t=50\,\rm s$, $t=100\,\rm s$ and $t=436\,\rm s$. The source and drain droplets are located on the $x$-axis with centers at the points $\bB{x}_1 = (-6.25,0)\, {\rm mm}$ (left position) and $\widetilde{\bB{x}}_1 = (6.25,0)\, {\rm mm}$ (right position), respectively, and both droplets have the same radius $1.5\,\rm mm$.
  The concentrations are given in $\rm kg/m^2$ units, while the height is in meters.}\label{fig:SIM1:Image1}
\end{figure*}

\subsection{Example 1: Configuration 1S1D}\label{sec:example1}
\begin{figure*}[!t]
\centering
 \includegraphics[scale=1]{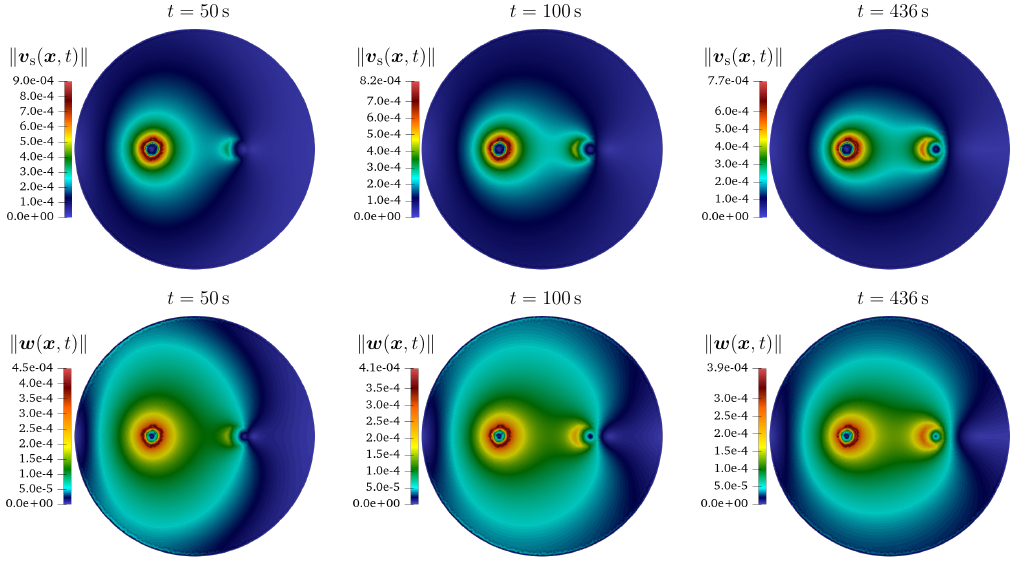}
  \caption{Example 1. Velocity magnitudes $\|\bB{v}_{\rm s}\|$ (first row) and  $\|\bB{w}\|$ (second row) at $t=50\,\rm s$, $t=100\,\rm s$ and $t=436\,\rm s$. The source and drain droplets are located on the $x$-axis with centers at the points $\bB{x}_1 = (-6.25,0)\, {\rm mm}$ (left position) and $\widetilde{\bB{x}}_1 = (6.25,0)\, {\rm mm}$ (right position), respectively, and both droplets have the same radius $1.5\,\rm mm$.
  The plots are presented in $\rm m/s$ units.}\label{fig:SIM1:Image2}
\end{figure*}

For the first example, we consider the first scenario presented in \citep{Nguindjel2024}. We simulate the case of one source droplet and one drain droplet both of radius $r_1 = \widetilde{r}_1 = 1.5\,\rm mm$, and located at
\begin{alignat*}{2}
\bB{x}_1 &= (-6.25,0)\, {\rm mm}\quad &&\text{(source)},\\
\widetilde{\bB{x}}_1 &= (6.25,0)\, {\rm mm}\quad &&\text{(drain)}.
\end{alignat*}
We set the constant rates $\kappa_{\rm b} = 0.01\,\rm s^{-1}$, $\kappa_{\rm d} = 1\,\rm s^{-1}$ and $\kappa_{\rm s} = 140\, \rm s^{-1}$.
The evolution of the surfactant spreading, height $h$ and bulk concentration $c$ are shown in Figure~\ref{fig:SIM1:Image1} at the time points $t = 50\,\rm s$, $t=100\,\rm s$ and $t=436\,\rm s$. Furthermore, the magnitude of the vectors $\bB{v}_{\rm s}$ and $\bB{w}$ at the same time points are presented in Figure~\ref{fig:SIM1:Image2}.
As expected, the spreading of surfactant begins radially until the solid particles meet the drain droplet or the boundary of the Petri dish. One of the effects of including the source term $\psi$ with the chosen parameter $\kappa_{\rm s} = 140\,\rm s^{-1}$ is that the concentration $\Gamma$ remains  constant on the inner circle (radius $r<\alpha r_1$), approximately equal to $\Gamma_{\rm m}$.
The depletion of surfactant produced by the drain drop generates a surface tension gradient towards the center of the drain. This effect can be observed in Figure~\ref{fig:SIM1:Image1} (first column) in view that the concentration assumes its minimum at the drain droplet. Besides, the bulk concentration, which initially is set to zero, increases proportionally within the domain. The height of the liquid film, shown in the third row of Figure~\ref{fig:SIM1:Image1}, exhibits the expansion wave produced by the source droplet, which raises from $2.5\times 10^{-7}\,\rm m$ to $h_0 = 2.6\times10^{-6}\,\rm m$ at its maximum point. In Figure~\ref{fig:SIM1:Image2} (first row), we observe that, at each time reported, the maximum spreading speed is given at the boundary of the source droplet, and that the velocity magnitude $\|\bB{v}_{\rm s}\|$ rapidly decreases to numbers in orders of magnitude between $0.1\,\rm mm/s$ and $0.3\,\rm mm/s$, which is the most prevalent range of values within the domain.
The magnitude of $\bB{w}$, on the other hand, is approximately half that of $\bB{v}_{\rm s}$, presenting a broader expansion in space as it evolves in time.

\begin{figure*}[!t]
\centering
 \includegraphics[scale=1]{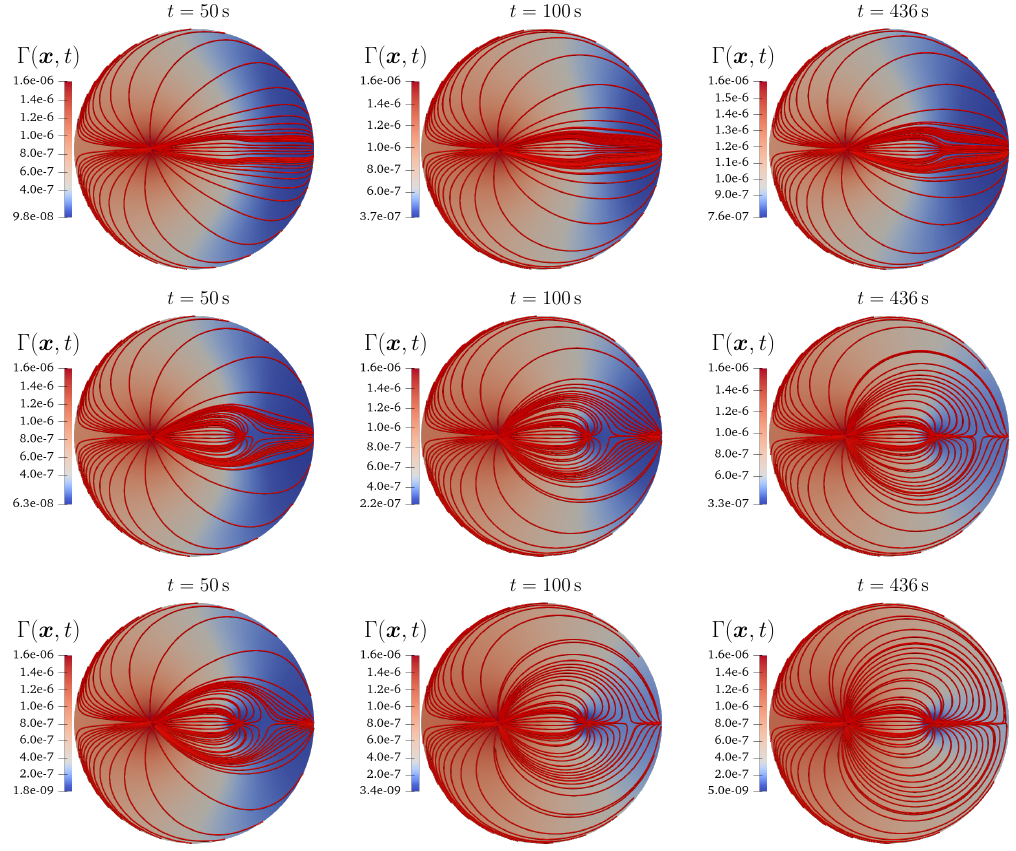}
 \caption{Example 1. Surfactant concentration $\Gamma$ including streamlines (red lines) computed with three values of $\kappa_{\rm d}$, $\kappa_{\rm d} = 0.1\,\rm s^{-1}$ (first row), $\kappa_{\rm d} = 1\,\rm s^{-1}$ (second row) and $\kappa_{\rm d} = 10\,\rm s^{-1}$ (third row) at $t=50\,\rm s$, $t=100\,\rm s$ and $t=436\,\rm s$. In the three cases, the source and drain droplets are located on the $x$-axis with centers at the points $\bB{x}_1 = (-6.25,0)\, {\rm mm}$ (left position) and $\widetilde{\bB{x}}_1 = (6.25,0)\, {\rm mm}$ (right position), respectively, and both droplets have the same radius $1.5\,\rm mm$.
  The plots are presented in $\rm kg/m^2$ units.}\label{fig:SIM1:Image3}
\end{figure*}

One of the advantages of solving the equations with $\bB{v}_{\rm s}$ as an additional unknown, is that this vector can be used to determine the streamlines which, in turn, correspond to the trajectories followed by the surfactant wires in \citep{Nguindjel2024}. To illustrate the streamlines and the effect that the drain droplet has on them, we simulate the process with six values of $\kappa_{\rm d}$ \eqref{eq:def:drain}, these are $\kappa_{\rm d} = 0.1,\, 0.5,\, 1,\, 2,\, 10\,\rm s^{-1}$. In Figure~\ref{fig:SIM1:Image3}, we show $\Gamma$ including a number of streamlines (red lines) for $\kappa_{\rm d} = 0.1\,\rm s^{-1}$ (first row), $\kappa_{\rm d}=1\,\rm s^{-1}$ (second row) and $\kappa_{\rm d} = 10\,\rm s^{-1}$ (third row) at the times $t=50\,\rm s$, $t=100\,\rm s$ and $t=436\,\rm s$. The streamlines are produced in the same conditions (number of lines, length and numerical method) for the three cases using the open-source software Paraview.
We observe that, as expected, an increase in $\kappa_{\rm d}$ produces an increase in the attraction of the streamlines.
This effect is more pronounced in the part behind the drain droplet, for $x>6.25\,\rm mm$, where in the third case the flow lines curve completely towards the drain at $t=436\,\rm s$.

\begin{figure*}
\centering
 \includegraphics[scale=1]{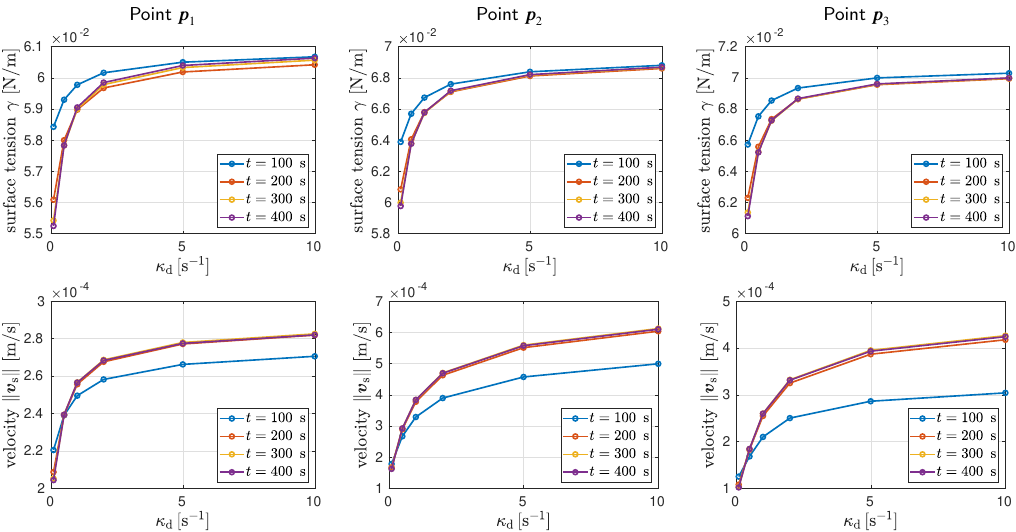}
\caption{Example 1. Surface tension $\gamma$ (first row) and magnitude of the velocity $\bB{v}_{\rm s}$ (second row) varying with respect to $\kappa_{\rm d}$ at four different times $t=100,200,300,400\,\rm s$ and three spatial locations.}\label{fig:SIM1:Image4}
\end{figure*}

The six cases are compared on Figure~\ref{fig:SIM1:Image4} at three spatial locations:
\begin{alignat*}{2}
 \bB{p}_1 &= (0,0)\,\rm mm&&\quad \text{(center-origin)},\\  
 \bB{p}_2 &= (4,0)\,\rm mm&&\quad \text{(left to drain)},\\ 
 \bB{p}_3 &= (6.25,-2.25)\,\rm mm &&\quad \text{(below drain)}. 
\end{alignat*}
At the three points analyzed, we observe from the first row in Figure~\ref{fig:SIM1:Image4} that the surface tension increases with respect to $\kappa_{\rm d}$ and respectively decreases (in a less pronounced way) from $t=100\,\rm s$ for each $\kappa_{\rm d}$ value. However, from $t=200\,\rm s$ to $t=400\,\rm s$ for $\bB{p}_1$ and $\bB{p}_2$, the curves tend to coincide. Although the values at $\bB{p}_1$ and $\bB{p}_2$ are comparable, the surface tension is higher in $\bB{p}_3$, and the smaller values are assumed in $\bB{p}_1$. For the speed of spreading, we can determine from the second row in Figure~\ref{fig:SIM1:Image4} that the magnitude of $\bB{v}_{\rm s}$ is increasing with respect to $\kappa_{\rm d}$ and time in the three chosen points, $\bB{p}_1$, $\bB{p}_2$ and $\bB{p}_3$.

\subsection{Example 2: Configuration 1S4D}\label{sec:example2}
\begin{figure*}[!t]
\centering
 \includegraphics[scale=1]{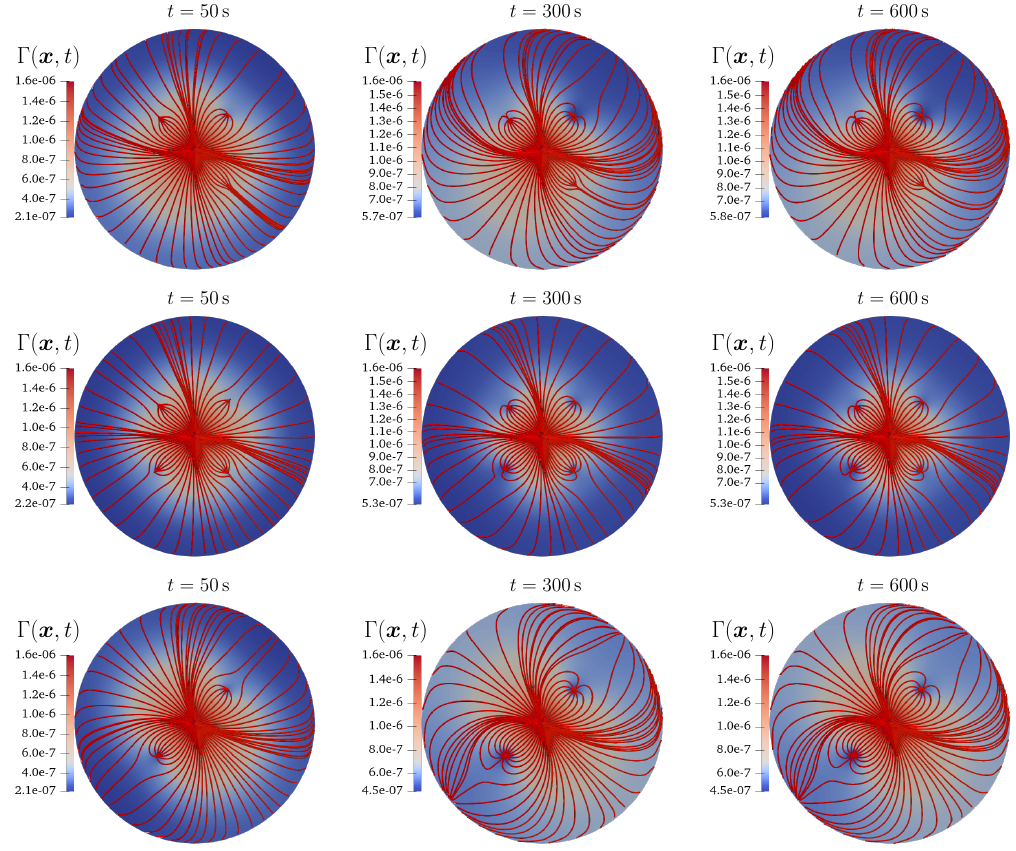}
 \caption{Example 2. Surfactant concentration $\Gamma$ including streamlines (red lines) for the three cases \textbf{C1} (first row), \textbf{C2} (second row) and \textbf{C3} (third row) at $t=50\,\rm s$, $t=300\,\rm s$ and $t=600\,\rm s$. The plots are in $\rm kg/m^2$ units.}\label{fig:SIM2:Image1}
\end{figure*}

We now simulate the second scenario considered in \citep{Nguindjel2024}, this is, the configuration of one source droplet and four drain droplets. The radius of the source droplet is $r_1 = 1.5\,\rm mm$, with center point at the origin $\bB{x}_1 = (0,0)\,\rm mm$. For the drain droplets, we set the same radius for all four droplets  $\widetilde{r}_1 = \widetilde{r}_2 = \widetilde{r}_3 = \widetilde{r}_4 = 1.0\,\rm mm$, and place their positions with the center points:
\begin{alignat*}{2}
\widetilde{\bB{x}}_1 & = (-5.0129,-4.5185)\,\rm mm &&\quad \text{(bottom-left)},\\
\widetilde{\bB{x}}_2 & = (\phantom{+}4.1870,-4.6296)\,\rm mm &&\quad \text{(bottom-right)},\\
\widetilde{\bB{x}}_3 & = (-4.5240,\phantom{+}3.8889)\,\rm mm &&\quad \text{(top-left)},\\
\widetilde{\bB{x}}_4 & = (\phantom{+}4.3703,\phantom{+}4.6296)\,\rm mm &&\quad \text{(top-right)}.
\end{alignat*}
We set the values of $\kappa_{\rm b}$ and $\kappa_{\rm s}$ as in~Example 1.
In this example, we consider individual constant rates $\kappa_{\rm d,1}$, $\kappa_{\rm d,2}$, $\kappa_{\rm d,3}$ and $\kappa_{\rm d,4}$ corresponding to $\kappa_{\rm d}$ for each drain droplet with indices $i=1,\dots,4$ according to the center points, respectively. Three cases are studied depending on the values of $\kappa_{\rm d}$:
\begin{align*}
 \textbf{C1: }\quad & \kappa_{\rm d,1}=0.001\,{\rm s^{-1}},\quad \kappa_{\rm d,2}= 0.5\,{\rm s^{-1}},\\
                    &\kappa_{\rm d,3}=1\,{\rm s^{-1}},\quad \kappa_{\rm d,4} = 1.5\,{\rm s^{-1}},\\[1ex]
 \textbf{C2: }\quad & \kappa_{\rm d,1}=\kappa_{\rm d,2}=\kappa_{\rm d,3}=\kappa_{\rm d,4} = 1\,{\rm s^{-1}},\\[1ex]
 \textbf{C3: }\quad & \kappa_{\rm d,1}=\kappa_{\rm d,4} = 0.1\,{\rm s^{-1}},\quad
                         \kappa_{\rm d,2}=\kappa_{\rm d,3} = 2\,{\rm s^{-1}}.
\end{align*}
The source surfactant $\Gamma$ including a number of streamlines is presented in Figure~\ref{fig:SIM2:Image1} for the three cases \textbf{C1} (first row),
\textbf{C2} (second row) and \textbf{C3} (third row) at the time points $t=50\,\rm s$, $t=300\,\rm s$ and $t=600\,\rm s$.
For the case of \textbf{C1}, it can be observed that the surfactant gets attracted to three of the drain droplets, except the one with the smallest value of $\kappa_{\rm d}$ (bottom-left drain droplet). Furthermore, the strength of attraction can be seen in the number of streamlines crossing through the drain, and the curvature of the lines on the back side of the droplet (region between the droplet and the boundary). In the case of \textbf{C2}, the drain droplets attract the surfactant with the same rate, and the solution becomes symmetric with respect to both axes at three times reported. In the third case \textbf{C3}, once again we observe that the two droplets with the smaller value of $\kappa_{\rm d}$ have practically no influence on flow lines, while the these lines get curved towards the drain droplets of higher values of $\kappa_{\rm d}$.

\subsection{Example 3: Configuration 2S1D (moving)}

\begin{figure*}[!t]
\centering
  \includegraphics[scale=1]{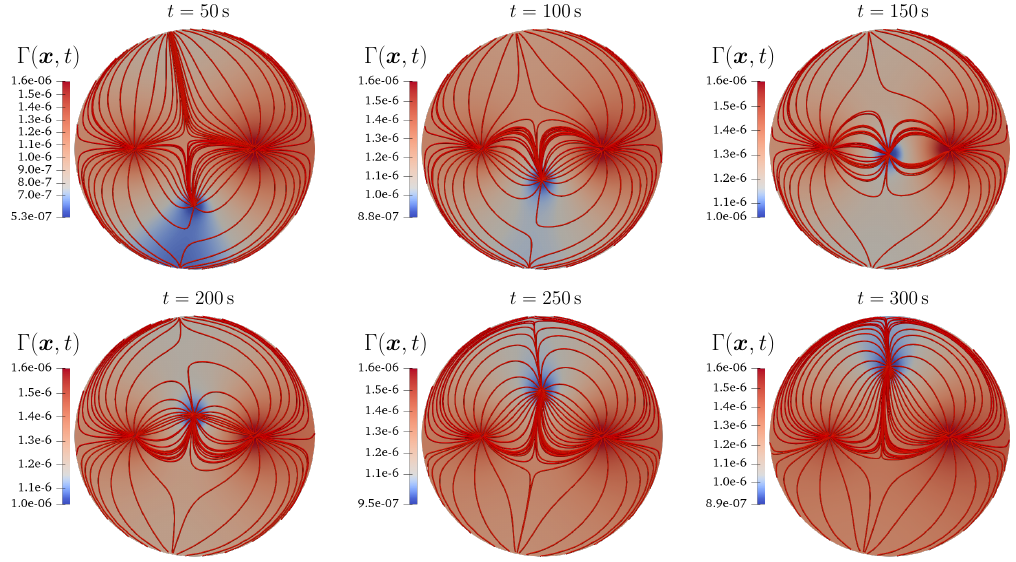}
 \caption{Example 3. Surfactant concentration $\Gamma$ including streamlines (red lines) at the times $t=50,100,\dots,300\,\rm s$. The drain droplet of constant radius $\widetilde{r}_1=1\,\rm mm$ characterized by its center point $\widetilde{\bB{x}}_1$ moves from $(0,-10.5)\,\rm mm$ at $t=0\,\rm s$ to $(0,10.5)\,\rm mm$ at $t=300\,\rm s$.
 The plots are in $\rm kg/m^2$ units.}\label{fig:SIM3:Image1}
\end{figure*}

For the final example, we simulate a drain droplet of radius $\widetilde{r}_1=1\,\rm mm$, moving along the vertical axis and between two source droplets in a configuration 2S1D. In this case, we consider two droplets with different radius $r_1= 1\,\rm mm$ and $r_2 = 2\,\rm m$, centered at the points
\begin{align*}
 \bB{x}_1 &= (-8.75,0)\,\rm mm\quad \text{(left source)},\\
 \bB{x}_2 &= (\phantom{+}8.75,0)\,\rm mm\quad \text{(right source)}.
\end{align*}
The rate of depletion at the drain droplet is set to $\kappa_{\rm d}=1\,\rm s^{-1}$, and the rate of surfactant release at the source droplet is set to $\kappa_{\rm s} = 120\,\rm s^{-1}$, $\kappa_{\rm b}$ is chosen as in Example 1 and 2. To produce the movement of the drain, we consider its center $\widetilde{\bB{x}}_1$ varying with respect to time as follows
\begin{align*}
 \hspace{-0.5cm}\widetilde{\bB{x}}_1(t):= (0,-10.5) + \dfrac{21\,t}{300}\,(0,1)\,{\rm mm},\quad 0\,{\rm s}\leq t\leq 300\,{\rm s},
\end{align*}
where $\widetilde{\bB{x}}_1(0) = (0,-10.5)\,\rm mm$ and $\widetilde{\bB{x}}_1(300) = (0,10.5)\,\rm mm$. The simulated surfactant concentration $\Gamma$, including a number of streamlines, is shown in Figure~\ref{fig:SIM3:Image1} for six different times $t=50,100,\dots,300\,\rm s$.
From the time sequence of $\Gamma$, it is clearly seen how the flow lines of both source droplets are curved towards the drainage as it moves upward in the vertical axis. In addition, we note that the surfactant concentration in the domain increases until $t=200\,\rm s$, but that after the drain droplet passes by, this concentration begins to decrease.

For $t=50\,\rm s$ and $t=100\,\rm s$, the difference in radius of both source droplets is clearly seen, because one of the rings of concentrations is larger than the other.
However, from $t=150\,\rm s$ onwards, this difference is difficult to see with the naked eye, and the main one is that the line to which the streamlines going towards the south pole of the domain converge, is slightly inclined.

\section{Conclusions}
We have proposed a two-dimensional, time-dependent model for the simulation of the spreading of surfactant droplets on liquid films containing drain droplets. The model equations developed here are based on the thin-film equations, which are formulated in accordance with the hypothesis of thin liquid film.
In the experiments shown by \citep{Winkens2022} and \citep{Nguindjel2024}, the height of the liquid film is about 15\% of the characteristic length $L$. However, the imposition of this assumption is based on the argument that surfactant dynamics takes place in the limit of the water-air interface leading to very small variations in height.
Also, we note that our model does not include the Marangoni counter flow under the air-water interface (experimentally observed to be in opposite direction to the Marangoni flow along the air-water interface, \cite{Winkens2022}). Nevertheless, the model captures the flow patterns along the air-water interface from source to drains.

Our model provides a variety of possibilities for simulating different scenarios.
As presented in the examples shown in Section~\ref{sec:simulations}, the model allows incorporating an arbitraty number of source and drain droplets, having the obvious limitation of the bounded domain and disjoint droplets. Among the adjustable parameters, we have the radius and depletion rate at the drain droplets, and depletion rate related to the bulk concentration. The initial value $h_0$ and constant $\varepsilon = h_0/L$ can also be modified, being the second constant to the one that is strongly related to the propagation speed of the surfactant. In addition, the model and numerical approximation framework, how it has beed presented, is not restrictive regarding initial conditions, the choice of the $\gamma$ function and reaction rate functions $\mathcal{R}_{\rm b}$ and $\mathcal{R}_{\rm d}$. The circular shape of the droplets is not a limitation, therefore the model can be extended to more general geometries.

From the examples shown in Section~\ref{sec:simulations}, we can conclude that, although we are not modeling the wire of filaments produced in the experiments presented in \citep{Nguindjel2024}, the velocity field $\bB{v}_{\rm s}$ allow us to compute flow lines which predict the curvature and direction of the filaments. Furthermore, one of the advantages of our approach is that we compute $\bB{v}_{\rm s}$ as an independent unknown, therefore we can determine the speed of propagation at each point of the domain and each time. Another benefit of our approach is that we can determine the height of the liquid $h$ and the speed of wave propagation produced at the surface of the liquid film related to $\bB{w}$.

There exist several ways to continue developing mathematical and physicochemical aspects of these types of systems. One of the main questions is related to physics behind the drain droplets. Several experiments \citep{Winkens2022,Winkens2023,Nguindjel2022,Nguindjel2024} suggest that drain droplets rotate and translate as entire identities. In this regard, an additional equation for the modeling of $\widetilde{\Gamma}$ as a function varying in space and time needs to be provided. On a first approach, if no variations in space are considered, that is $\widetilde{\Gamma}=\widetilde{\Gamma}(t)$, the following ordinary differential equation would be sufficient:
\begin{align*}
 \dfrac{\partial\widetilde{\Gamma}}{\partial t} = \mathcal{R}_{\rm d}(\Gamma,\widetilde{\Gamma},c)\qquad\bB{x}\in \Omega,
\end{align*}
with $\mathcal{R}_{\rm d}$ as a reaction term depending on the concentration of $\widetilde{\Gamma}$, $\Gamma$ and $c$.
For the coupling of an equation of this type to make sense in our system, the reaction term $\mathcal{R}_{\rm s}$ would have to be reformulated effectively depending on the concentration $\widetilde{\Gamma}$.
A more elaborated formulation accounting the translation of the drain droplet could be given by the following convection-reaction equation:
\begin{align*}
 \dfrac{\partial\widetilde{\Gamma}}{\partial t}  + \divr\bigl(\widetilde{\Gamma}\bB{u}\bigr) =  \mathcal{R}_{\rm d}(\Gamma,\widetilde{\Gamma},c)\qquad\bB{x}\in \Omega,
\end{align*}
where $\bB{u}$ being the transport velocity related to the drain droplet. The main challenge in obtaining such an equation for describing $\widetilde{\Gamma}$, is to find a correct way of defining the velocity field $\bB{u}$.
In the event that the drain droplet does not deform in space while moving, the velocity field $\bB{u}$ could be considered constant within the droplet, varying only in time. The main inconvenience would then be the rotation of the droplets. In conclusion, further studies regarding the forces acting on the drain droplet and $\widetilde{\Gamma}$-dependent reaction terms need to be done.

In our model, the source droplet is treated in the same way as the surfactant being spread, and the source droplet remains as an entity due to the term $\psi$ in Equation~\eqref{eq:mainevol:b}, which keeps the concentration approximately constant at $\Gamma_{\rm m}$. This allows both the source droplet and the surfactant concentration to be treated with a single equation. An alternative model would be to consider the source droplet as an additional solid phase on top of the surface, which releases surfactant to the water-air surface from its boundary.

The thin-film assumption, which translates into a small value of $h_0$, does not affect the overall process and the results obtained are in accordance with the experiments. However, a different approach can be addressed by considering a shallow water or Saint-Venant formulation or the fully three-dimensional Navier-Stokes equations \citep{Careaga2024, Osores2020}. Regarding the numerical scheme proposed based on mixed finite elements, this allows the variables to be calculated as a single vector making use of the classical Lagrange finite element spaces. The implicit approximation of the time derivative enables the calculations to be carried out without requiring a severe restriction on the time step.
\section{Acknowledgements}
This work has been partially supported by the IRP Voucher Project no.~6201479 from Radboud University. J.C. is supported
by ANID Chile through the Fondecyt project no.~3230553.

\bibliographystyle{elsarticle-harv}

\end{document}